\documentclass[preprint,showpacs,preprintnumbers,nofootinbib,amsmath,amssymb]{revtex4}

\usepackage{graphicx}
\usepackage{dcolumn}
\usepackage{bm}

\newcommand{\be}{\begin{equation}}
\newcommand{\ee}{\end{equation}}
\newcommand{\bdm}{\begin{displaymath}}
\newcommand{\edm}{\end{displaymath}}
\newcommand{\bea}{\begin{eqnarray}}
\newcommand{\eea}{\end{eqnarray}}

\newcommand{\bht}{\hat{\bf t}}

\newcommand{\llp}{\frac{L}{l_{p}}}

\newcommand{\lpa}{l_{p}}

\newcommand{\newysq}{y^2}
\newcommand{\newmysq}{(1 - y^2)}
\newcommand{\yfour}{y^4}

\begin{document}

\title{Distribution Functions, Loop Formation Probabilities and
Force-Extension Relations in a Model for Short Double-Stranded DNA Molecules}

\author{P. Ranjith$^1$\footnote{E-mail: ranjith@physics.iitm.ac.in},
P. B. Sunil Kumar$^1$\footnote{E-mail: sunil@physics.iitm.ac.in} and 
Gautam I. Menon$^2$\footnote {E-Mail: menon@imsc.res.in}}

\affiliation{$^1$Department of Physics, Indian Institute of Technology Madras,
Chennai 600 036, INDIA \\
$^2$The Institute of Mathematical Sciences, C.I.T Campus,
Taramani, Chennai 600 113, INDIA}

\date{\today}

\begin{abstract}
We obtain, using transfer matrix methods, the distribution
function $P(R)$ of the end-to-end distance, the loop
formation probability and force-extension relations
in a model for short double-stranded DNA molecules.
Accounting for the appearance of ``bubbles'', localized
regions of enhanced flexibility associated with the
opening of a few base pairs of double-stranded DNA in
thermal equilibrium, leads to dramatic changes in $P(R)$
and unusual force-extension curves.  An analytic formula
for the loop formation probability in the presence of
bubbles is proposed.  For short {\em heterogeneous} chains,
we demonstrate a strong dependence of loop formation
probabilities on sequence, as seen in recent experiments.
\end{abstract}

\maketitle

The physics of bending and loop formation in DNA
is key to a variety of regulatory processes within
the cell.  Loop formation in DNA is believed to be
central to enhancer action, while the compactness of
DNA packaging within the nucleosome necessitates the
bending of DNA over length scales of a few tens of base
pairs~\cite{kadonaga-98,alberts}.  In a broader context, 
the modelling of bending and looping in biopolymers
at length scales over which intrinsic stiffness plays a 
dominant role is a problem of general relevance.

Worm-like chain (WLC) models of DNA elasticity incorporate
semi-flexibility, describing the chain in terms of a
persistence length $l_p$, the length-scale at which tangent
vectors to the polymer are decorrelated~\cite{doi-ed}.
On scales smaller than $l_p$, bending energy dominates and
the chain is relatively stiff.  It has conventionally been
assumed that the relatively large energy required to bend
short double stranded (ds) DNA of length $L \sim  100$~
base-pairs (bp) into loops necessitates the intervention of 
DNA-binding proteins.  Recent DNA cyclization experiments of Cloutier
and Widom (CW) which study relatively {\em small, isolated}~
dsDNA sequences question this assumption~\cite{cw}. In
the CW experiments, DNA molecules that are 94 bp in
length, comparable to sharply looped DNAs in {\it
vivo}, spontaneously cyclize with a large probability.
Theories of DNA elasticity based on the homogeneous WLC model
predict cyclization probabilities three to four orders of magnitude smaller
than those obtained experimentally~\cite{shimada-yamakawa},
indicating, in the words of Cloutier and Widom, ``a need
for new theories of DNA bending''~\cite{cw}.

The problem posed by these experiments has stimulated
much recent work on modelling loop formation in short
DNA molecules\cite{yan-marko,wiggins-nelson-kink}.
An attractive explanation for this discrepancy
is the existence of ``bubbles'', localized
regions of large flexibility induced by the
opening of a few base pairs of dsDNA in thermal
equilibrium~\cite{yan-marko}.  (Alternatively,
it has been argued that non-linear elastic effects
relevant at high curvature might induce a kinking
transition\cite{wiggins-nelson-kink}.)  Bubbles (or
kinks) are
argued to greatly increase the flexibility of
the WLC in their vicinity, thereby enhancing the
cyclization probability\cite{yan-marko,wiggins-nelson-kink}.
Recent transfer-matrix
based calculations implement this idea, but restrict
themselves to the computation of the cyclization
probability\cite{yan-marko}.

In this Letter we report the first calculation of the distribution function
of the end-to-end distance $P(R)$ in a model for
short dsDNA fragments with equilibrium bubbles. We
propose an analytic formula describing the loop
formation probability density $P(0)$ in such fragments and
compare this formula with results from numerical
calculations.  Varying the chemical potential for
bubbles leads to behaviour which interpolates between
fully flexible, semiflexible and rigid rod limits,
resulting in a variety of non-trivial force-extension
curves. Simple extensions of our model which simulate
DNA heterogeneity indicate that such heterogeneities
can be important determinants of the cyclization
probability in small chains.

We use the well-known connection of the 
WLC model, with hamiltonian $H_{WLC}= \kappa/2 \int_0^L
ds \left( \partial {\bf t}(s)/\partial s\right)^2$
constrained by $|{\bf t}^2(s)|=1$, to the Heisenberg spin
model ~\cite{fisher, bensimon}.  Here ${\bf t}(s) \equiv
\partial {\bf r}(s)/\partial s$ is the unit tangent vector
and  $\kappa$ is the bending stiffness. The persistence
length is $l_p = \beta\kappa$ with $\beta = 1/k_{B}T$.  Mapping the
continuum model to the discrete one requires that a minimum
coarse-graining length scale be specified. We fix this
to be $b=1$ nm (3 bp's). (This
also represents the scale at which the smallest bubbles
appear.) Thus, a 150 bp chain is represented by an L=50
site spin model. In what follows, both $L$ and $l_p$ are
dimensionless numbers representing the physical chain
length and persistence length in units of this basic
scale. The bending stiffness and the 
coupling constant in the spin model are related through
$J=\kappa /b$. In 
the WLC model, the distribution function of the end-to-end vector $P({\bf
R})$ characterizes the conformations of the polymer; it depends only on the
ratio $L/l_p$.  Rotational invariance imposes $P({\bf
R})=P(|{\bf R}|)=P(R)$.

The energetics of ds DNA in the presence of bubbles
in equilibrium is modelled {\it via} the following 
hamiltonian\cite{yan-marko}
\be
H = \sum_{i=1}^{N-1} \left[\tau_i J_s(1-\bht_i \cdot \bht_{i+1}) 
+  (1-\tau_i)J_d  (1-\bht_i \cdot \bht_{i+1}) \right] 
- {\mathrm \mu}\sum_{i=1}^{N-1}(1 - \tau_{\mathrm i}).
\label{Hamiltonian}
\ee
Here $\tau_i$ is a local variable which specifies whether a
given bond forms part of a bubble or not: $\tau_i = 1$ if
the site $i$ is a part of the bubble, otherwise $\tau_i
= 0$.  A chemical potential $\mu$ controls the energetics of the
$\tau_i$ variable and hence the number of bubbles\cite{lavery}. 
The bending stiffness of double-stranded
and bubble regions map to coupling constants 
$J_d$ and $J_s$  respectively.

The distribution function $P({\bf R})$ of the end-to-end vector
${\bf R}=\sum_{i=1}^N \hat{{\bf t}}_i$ is 
\be
P({\bf R})= \mathcal{N} \int d{\bf t}_1 \ldots \int d{\bf t}_N \sum_{\tau_1=0}^1... \sum_{\tau_{N-1}=0}^{1} e^{-\beta H}
\delta(\sum_i^{N-1} {\bht}_i -\bf{R}),
\ee
where ${\mathcal N}$ fixes $\int d{\bf R} P(R) =1$.
With one end of the polymer at ${\bf R}=0$,
the probability $p(z)$ for the other to be in a
given $z$-plane is related to $P({\bf R})$ through
$p(z)=\int d{\bf R} P(R) \delta(R_3 - z)$,
where ${\bf R}= R_1 \hat{e}_x+ R_2 \hat{e}_y+ R_3 \hat{e}_z$.
Defining a generating function $\tilde{p}(f)$~\cite{sam} through
$\tilde{p}(f)=\int_{-L}^L dz~\exp(fz) p(z)$,
yields, upon substituting definitions of $p(z)$ and $P(R)$,  the
relation $\tilde{p}(f)=Z(f)/Z(f=0)$, where 
$Z(f)= \int d{\bf t}_1 \ldots \int d{\bf t}_N \sum_{\tau_1=0}^1... 
\sum_{\tau_{N-1}=0}^{1} \exp{(-\beta H +
f \sum_{i=1}^N \hat{t}_i^z))}$.
After performing the $\tau_i$ summation, $Z(f)$ can be 
computed using transfer matrices~\cite{blume}.
Once $Z(f)$ is known, $\tilde{p}(f)$ and $p(z)$ can be 
computed. $P(R)$ is then finally obtained, exploiting the
tomographic methods outlined in Ref.~\cite{sam} and
symmetry arguments, from $P(R)=(-1/2 \pi z)dp(z)/dz|_{z=R}$.
We can either allow 
the tangent vectors at the two ends to fluctuate independently,
corresponding to free boundary conditions on the tangent vectors,
or require that they be equal, a boundary
condition believed to be appropriate to the CW 
experiments.

We benchmark our methods by computing $P(R)$ for a
homogeneous semiflexible polymer in the absence of bubbles,
({\it i.e.} $\mu=\infty$), at various $L/l_p$ values.
Our approach yields results which are fully consistent
with previous work, reproducing known answers from
simulations and analytic work
\cite{marko-siggia,frey}. In addition, for $L/l_p = 3.85$,
we recover the double hump feature in the distribution
function reported recently~\cite{abhishek,sam}.  We have
also computed the loop formation probability density 
$P(0)$~(defined as $P(R = 0)$) for the entire range of 
$L/l_p$ values, obtaining
results which agree well with the Shimada-Yamakawa
formula\cite{shimada-yamakawa}.

\begin{figure}[!h]
\includegraphics[width=8.8cm]{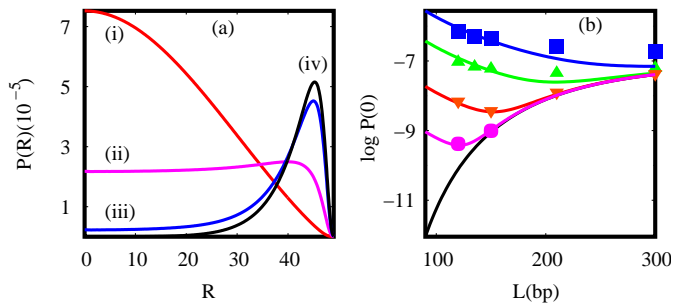}
\caption{\label{fig-pofr_bubble_L50}
(a) $P(R)$ calculated for a $150$ bp DNA fragment with 
$\beta J_d=50$ and $\beta J_s=1$ for (i)$\beta\mu=6$, (ii) $\beta\mu=7$ and 
(iii) $\beta\mu=9$. The curve (iv) is $P(R)$ from the 
WLC calculations($\beta \mu=\infty$). (b) $P(0)$
from the transfer matrix calculation (points) compared to the 
predictions of the aproximate formula of Eq. \ref{formula_cl} (lines).
The curves from top to bottom are for 
$\beta\mu = 10, 12, 15 ,18 $ and $\infty$. }
\end{figure}

To incorporate bubble formation, we fix $\beta J_d=50$ for the
double stranded region and $\beta J_s=1$ for the bubble region,
consistent with measures of the persistence length
\cite{notewell}.  Our results for $L=50$
are plotted in Fig.~\ref{fig-pofr_bubble_L50}(a) for $\beta \mu=6, 7$
and $9$. (The free energy cost in units of temperature $\beta\mu$
to form a $3$bp bubble can be estimated to lie between  $6$ and
$15$ depending on the sequence~\cite{yan-marko}).
For $\beta \mu=6$ the distribution function peaks at
$R=0$. Remarkably, for  $\beta \mu=9$, the distribution
function alters completely, with the peak shifting to
near $R=L$. For intermediate values of $\mu$, the peak
of $P(R)$ interpolates between $R=0$ and $R\simeq L$.
For $L/l_p=1$, $P(R)$ exhibits a double
hump feature when $\beta \mu=6.9$, as described earlier for
semiflexible chains in the regime in which $L/l_p \sim
3.85$. 

The loop formation probability density
$P(0)$ is experimentally accessible.
Fig.~\ref{fig-pofr_bubble_L50}(b) shows $P(0)$
(symbols) at different values
of $\mu$, as $L$ is varied. The boundary condition
imposed allows the tangent vectors at the chain ends
to fluctuate independently.  The lines through the
data points are predictions of the analytic theory
described below. At large $\mu$, the results asymptote
to those obtained for $\mu \rightarrow \infty$. As
$\mu$ is decreased from infinity, bubbles are favoured
and $P(0)$ increases sharply
at small $L$.

\begin{figure}[!h]
\includegraphics[width=8.5cm]{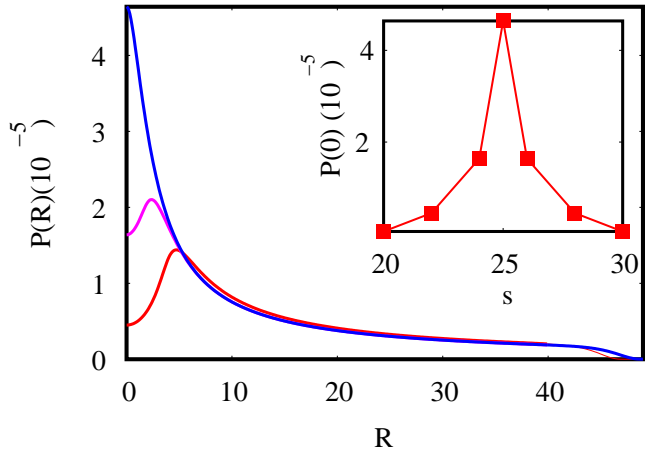}
\caption{\label{fig-1bubble} $P({\bf R})$ for 
homogeneous dsDNA of length $L=50$ computed 
with a single bubble placed  at L/2 (top curve), 
$L/2 \pm 1$ (middle curve) and  $L/2 \pm 3 $ (bottom curve).
The inset shows $P(0)$
as a function of bubble position, illustrating how this
quantity peaks sharply when the bubble is placed at the centre.
}
\end{figure}

We have specialized our calculations to allow for the
insertion of a single bubble at arbitrary points along
the chain. This allows us to check for the optimal bubble 
location.  We compute $P(R)$ for the
case in which a single bubble is placed at the centre
($L/2)$ as well as at positions which deviate from the
central position by one and two sites on both sides,
as shown in Fig.~\ref{fig-1bubble}. At the parameter
values $\beta J_d=L=50$, the distribution function is peaked sharply near
$L$. Allowing for a single bubble at the centre transforms
$P(R)$ completely, shifting the peak to $R=0$.  The peak
of the distribution function moves away from $R=0$, as
the bubble position moves off-centre, with $P(R)$ peaking
near $R=L$ as the bubble position shifts to the chain end.
We plot $P(0)$ as a function of the bubble
position in the inset to Fig~\ref{fig-1bubble}.
As the bubble position moves 3 units away from the
center, $P(0)$ drops by one order of magnitude. This
peak at $L/2$  becomes sharper for small $L$,
implying that the principal contribution to the loop
forming probability density for short chains comes from bubbles
positioned at the chain center.

This observation justifies a simple analytic approach
to $P(R)$ and the loop formation probability density:
The distribution function $P(R)$ for short DNA molecules
($L \ll l_p$, so they may be assumed to be rigid to a first
approximation) with one bubble in the middle can be
represented in terms of two infinitely rigid rods of length $L/2$
connected with a flexible hinge.  In the limit where the
bending energy cost at the hinge is zero, the probability
distribution function is that of a random walk with two
steps of equal size $L/2$.  This yields
$P(R) =1/(2 \pi L^2 R)$,
a behavior similar to that obtained from the transfer matrix 
calculation; see the top curve of Fig.~\ref{fig-1bubble} (main
panel).

Analytic results beyond the assumption of
rigid rod behaviour can be derived for the loop
formation probability density at large $\mu$, at which
the effects of a single bubble dominate. We have found that
the distribution function of the end-to-end vector
$P({\bf R},L,\lpa)$ for a semiflexible chain of length $L$ with
a persistence length $\lpa$ {\em in a regime relevant to
the DNA molecules in the CW experiment}, is dominated by
the weighted sum of two terms. The first term,
$P_0(\lpa,L)$, is the contribution in the absence of
bubbles, while the second, $P_1(\lpa,L)$, reflects the
contribution from a  single bubble placed at the 
centre of the chain. Thus, 
$P({\bf R},L,\lpa)$ in the limit ${\bf R} \rightarrow 0$, the 
loop formation probability density, is then
\be
P(0) \simeq \frac{1}{1 + 2\lpa e^{-\beta\mu}}\left [P_0(\lpa,L)  +
2\lpa e^{-\beta \mu}P_1(\lpa,L)\right ]. 
\ee
(The factors of $2\lpa e^{-\beta\mu}$ are fixed by
the normalization appropriate to the calculation of the
loop formation probability density from the spin model.)
The Shimada-Yamakawa theory of the 
cyclization of semiflexible polymers provides quantitative
estimates of $P_0(\lpa,L)$\cite{shimada-yamakawa} in the limit
relevant to the Cloutier-Widom experiments.

The term $P_1(\lpa,L)$ represents the loop forming probability
density of a chain of length $L$ with a flexible hinge in the center.
We exploit an accurate variational expression for
the end-to-end distribution function of a semiflexible chain
of length $L$~\cite{thirumalai-ha}:
$P({\bf R},L) = \frac{C(g)}{L^3}\left [ \frac{1}{1 - (R/L)^2}\right ]^{9/2}
\exp \left [\frac{g}{1 - (R/L)^2} \right ]$
with $g = 9L/8\lpa$, and obtain $P(0)$ from
an integral over the product of $P({\bf R},L)$'s of the form
$\int d{\bf R} P({\bf R},L/2)P(-{\bf R},L/2)$.  
This leads to  
$P_1(\lpa,L)=\left(\frac{32 \pi C^{2}(g/2)}{L^3}\right) \int_0^1 dx e^{f(x)}$,
where $x=2R/L$ and $f(x) = 2 \ln x - 9\ln(1-x^2)-g/(1-x^2)$,
and $C(g/2)$ is a normalization factor~\cite{thirumalai-ha}. The integral is
first approximated by saddle point techniques. We then
multiply the resulting analytic expression by a further
constant factor of 2.273, to match with results 
obtained through numerical integration. This procedure
yields near perfect agreement with numerics in the range 
$0 < L/l_p < 3$; a range over which the integral varies across 
12 orders of magnitude.  Further specializing the resulting expression
to the $L/l_p$ regime explored in the CW experiments yields
a single formula for $P(0)$,
involving $\beta \mu$ and $\lpa$:
\bea
P(0) &=& \frac{1}{1 + 2 \lpa e^{-\beta\mu}}
\left[ 112.04 \frac{\lpa^{2}}{L^5} e^{-14.055 \frac{\lpa}{L} 
+ 0.246 \frac{L}{\lpa}} \right. \nonumber \\
&+& \frac{18.31}{\pi^{\frac{3}{2}}\lpa^3}~(2\lpa e^{-\mu})
\left. \frac{ \newysq {\newmysq}^{\frac{-15}{2}} 
(1+ \frac{16 \lpa}{3 L} + \frac{320 \lpa^2}{27 L^2})^{-2}
e^{\frac{-9 L}{8 \lpa}\left(\frac{\newysq}{1-y^2}\right)}
}
{\left(36 \yfour - 
36\newysq + \frac{27}{4} \newysq\llp + 
\frac{9 L}{4 \lpa}\right)^{\frac{1}{2} }}  
\right],
\label{formula_cl}
\eea
where $y^2 = 1.037 - 0.1667L/\lpa$.
This formula is compared to the results of the
transfer matrix calculations in Fig.~\ref{fig-pofr_bubble_L50}.
The agreement is satisfactory, particularly at large $\mu$, where
the ``single bubble'' approximation is expected to be accurate.

\begin{figure}[!h]
\includegraphics[width=8.5cm]{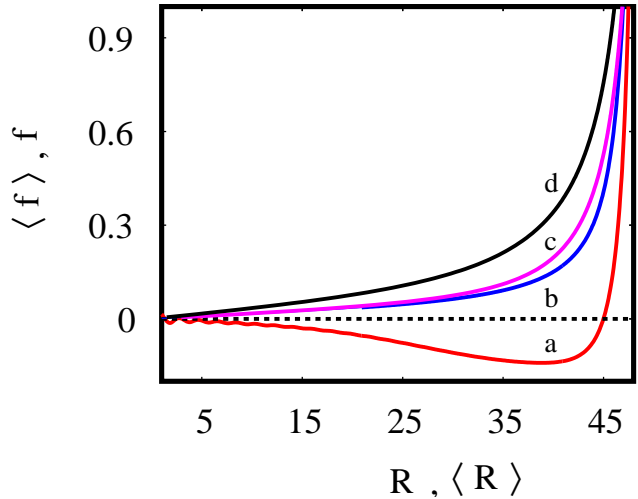}
\caption{\label{fe-L50}Force extension relation for
homogeneous dsDNA of length $L=50$ at parameter values 
$\beta J = 50$ with $\beta \mu=6$ and $\beta \mu=9$. The curves 
(a) and (b) are
computed in the fixed-extension ensemble whereas the
curves (c) and (d) are computed in the fixed force
ensemble.}
\end{figure}

Variations in $P(R)$ as a consequence of bubble formation
should be reflected in experimentally measurable force
extension (FE) relations. Experiments typically measure
the average force required to maintain the two ends of
the polymer at a fixed separation $R$. This average force
is $\langle f \rangle = -\partial \log{(P(R))}/\partial
R$. Since our calculations access $P(R)$
directly, we can calculate this average force as a
function of extension in all the cases discussed earlier.
We find that the FE curves are strongly
ensemble dependent for small chains -- FE relations in the fixed force
ensemble  are {\em always} monotonic\cite{wiggins-nelson-kink}, 
while non-monotonic
relations can be obtained in the fixed extension ensemble.  
Such non-monotonicity
disappears in the $L \rightarrow \infty$ limit, where FE
curves calculated in both ensembles coincide, as expected.
Fig.~\ref{fe-L50} shows plots for homogeneous chains
setting $\beta J = 50$ and $\beta \mu = 6$ and $9$, of
$\langle f\rangle$ {\it vs.} R, in the constant extension
ensemble (plots (b) and (a), respectively), as well as in the constant
force ensemble, in which $\langle R \rangle$ {\it vs.} f
is calculated (see plots (c) and (d)), illustrating these
conclusions. 

We have also investigated the effects of sequence
heterogeneity on the loop formation probability
density and $P(R)$. The energy to break paired
bases is strongly sequence dependent, and
alters the local value of $J$. (It is known
that A-T bonds are more easily broken than G-C
bonds~\cite{alberts,libchaber-bubble}.)  Intuitively,
regions of reduced bending rigidity should play a role
similar to that played by bubbles, reducing the energy
required to bend the chain at specific locations. We
have experimented with 50:50 mixtures of bonds
with strength $\beta J=40$ (weak) and $\beta J=60$
(strong) in a system of size $L=100$ and arranged in
the following way: (a) two equal stretches of strong
bonds at each end, separated by 50 weak bonds in
the central region (b) two equal stretches of weak
bonds at each end, separated by 50 strong bonds
in the central region  and (c) a typical random
sequence formed by laying weak and strong bonds down
at random, subject to the constraint that they are
equal in number.  We have checked that in the limit
of large chains, the results for a typical random
sequence of weak and strong bonds are close to those
obtained from the homogeneous case with $\beta J=50$.

Our results are the following: as is intuitively
clear, the case in which the stretch of weak bonds is
placed in the centre, case (a), yields the largest
values for $P(0)$ while the case
in which strong bonds populate the centre, case(b),
yields the smallest. The random chain result, case
(c), is close to the result for the homogeneous case.
The variation in $P(0)$
spans a {\em full} order of magnitude or more for
short chains at these parameter values, indicating
the importance of sequence heterogeneity for loop
formation.

In conclusion, we have computed a wide range of
physical properties of a simple model for short dsDNA
molecules which incorporates the presence of bubbles
in equilibrium.
We point out that several physical properties of
short dsDNA molecules, as reflected in $P(R)$,
are strongly affected by
bubble formation.  Our model is easily extended to
account for sequence heterogeneity.  The unusual
force-extension relations in diverse ensembles
exhibited here may have implications for loop
formation in {\it vivo}.

We thank M. Rao and R. Siddharthan for discussions and
J. Widom for sending us a copy of Ref.\cite{cw}. PR
acknowledges discussions with J. Marko and
S. Sankararaman. This work was partially supported by the
CSIR (India).

\end{document}